\documentclass[twoside]{LCWS11}
\usepackage[latin1]{inputenc}
\usepackage[dvips]{graphicx,epsfig,color}
\usepackage{wrapfig,rotating}
\usepackage{amssymb,amsmath,array}
\usepackage{cite}
\begin{document}
\title{
Beam-induced backgrounds in the CLIC 3 TeV CM energy interaction region} 
\author{B. Dalena$^{1,}$$^2$, J. Esberg$^{2,}$$^3$ and D. Schulte$^2$
\vspace{.3cm} \\
1- CEA/SACLAY, DSM/Irfu/SACM - 91191 Gif-sur-Yvette - France \\
2- CERN - Geneva - Switzerland \\
3- Department of Physics and Astronomy, Aarhus University, Denmark \\
}

\maketitle

\begin{abstract}
Luminosity spectrum and accelerator background levels strongly influence the 
experimental conditions and have an important impact on detector design.
The expected rates of the main beam-beam products at CLIC 3 TeV CM energy, 
taking into account for machine imperfections, are computed. 
Among the other machine-induced background the photon fans from the Incoherent
Synchrotron Radiation (ISR) photons emitted in the final doublet are
evaluated.    
\end{abstract}

\section{Introduction}

In the design of the CLIC interaction region the background levels
need to be carefully taken into account, since their rates are expected
to be high because of the high energy and high luminosity foreseen.
Two main sources of background can be identified: those coming from the
beam interactions before and after the collision point, the so called machine
backgrounds, and those arising from beam-beam effects, so called beam-beam
background. In this paper we review the main beam-beam products in order to
give an upper limit to their expected rates, the impact on the
luminosity spectrum is also discussed. The distribution of
their expected energy and angle are shown. Furthermore we discuss the 
impact of ISR photons coming from the final doublet on the CLIC interaction 
region. 
\begin{wraptable}{l}{0.5\columnwidth}\vspace{-0.2cm}
\centerline{\begin{tabular}{|l|c|c|}
\hline
Total Luminosity  & [10$^{34}$cm$^{-2}$s$^{-1}$] & 5.9 \\
Peak Luminosity  & [10$^{34}$cm$^{-2}$s$^{-1}$] & 2.4  \\
repetition freq. & [Hz] & 50 \\
bunches/train &  & 312 \\
intra-bunch dist.  & [ns] & 0.5 \\\hline
particles/bunch & [10$^{10}$] & 0.372 \\
bunch length & [$\mu$m] & 44 \\
emittances H/V & [nm]/[nm] & 660/20 \\
beam sizes & [nm]/[nm] & 45/1 \\
\hline
\end{tabular}}\vspace{-0.2cm}
\caption{CLIC parameters taking into account machine imperfections.}
\label{tab:1}
\end{wraptable}\vspace{-0.3cm}

\begin{figure}\vspace{-0.3cm}
\centerline{\includegraphics[width=0.78\columnwidth]{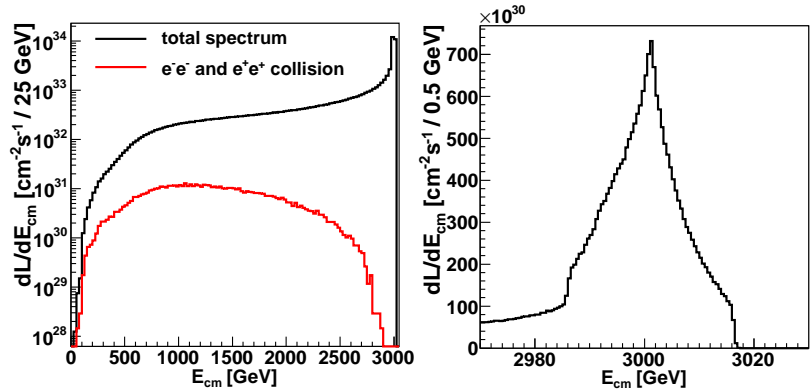}}\vspace{-0.2cm}
\caption{Luminosity spectrum for nominal CLIC 3 TeV CM energy parameters.}\label{Fig:1}
\end{figure}

In order to achieve the required luminosity the two beams at the future 
linear colliders are focused to very small sizes, see Table~\ref{tab:1}. 
In electron-positron collisions the electromagnetic field of each bunch will 
focus the other, leading to an enhancement of total luminosity (so-called
Pinch effect). At the same time due to the strong bending of their trajectory,
the beam particles emit high-energy photons (called beamstrahlung photons),
which smear the peak of the luminosity spectrum, as shown in 
Fig.~\ref{Fig:1}. 

In addition to beamstrahlung photons also QED and QCD 
backgrounds are produced during collision. The relevant processes are: coherent
pair production, incoherent pair production and $\gamma\gamma \to$ hadrons 
events. They are briefly described in the next section. The pairs produced in
the coherent processes can contribute to luminosity $\sim$4$\%$ of the 
total luminosity comes from these pairs mainly in the low energy tail of the 
spectrum. They can also create collisions
where an electron, from a coherent pair produced in the positron beam, collides
with the electron beam (and vice versa for a positron). The contribution
of these type of collisions to the luminosity is $\sim$1$\%$, which correspond
to the red line showed in Fig.~\ref{Fig:1}.\\

\vspace{-0.3cm}

\section{Beam-Beam backgrounds at 3 TeV CM energy}

The beam-beam backgrounds rates are computed using the GUINEA-PIG 
code~\cite{gp}. In the simulations we use realistic bunch shapes 
coming from the full tracking of the two beams in the LINAC and 
the BDS systems toward the Interaction Point (IP). 
For this purpose the C++ version of the code~\cite{gp++} has been extensively 
reviewed and further developed. 
The beam-beam effects and processes that can be studied by GUINEA-PIG are: 
emission of beamstrahlung photons, coherent
processes such as creation of pairs particles in the strong electromagnetic 
field of the two bunches, and incoherent processes such as incoherent pairs 
creation and hadronic events. Other QED processes such as Bhabhas
can be simulated as well. 

\begin{figure}[b]
\begin{minipage}[b]{7.1cm}
\centerline{\includegraphics[width=0.75\columnwidth]{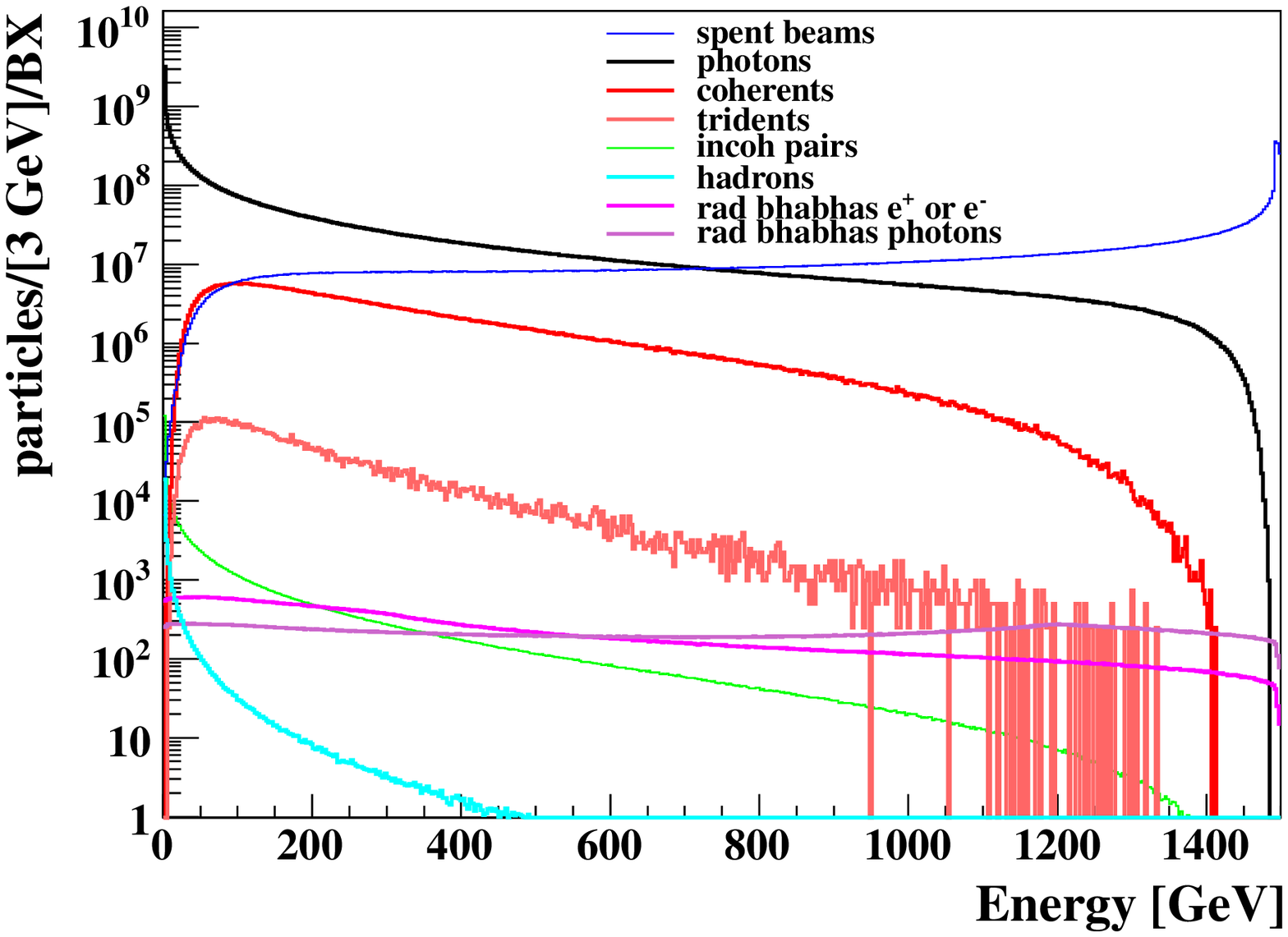}}\vspace{-0.2cm}
		
\end{minipage}
\begin{minipage}[b]{7.1cm}
\centerline{\includegraphics[width=0.81\columnwidth]{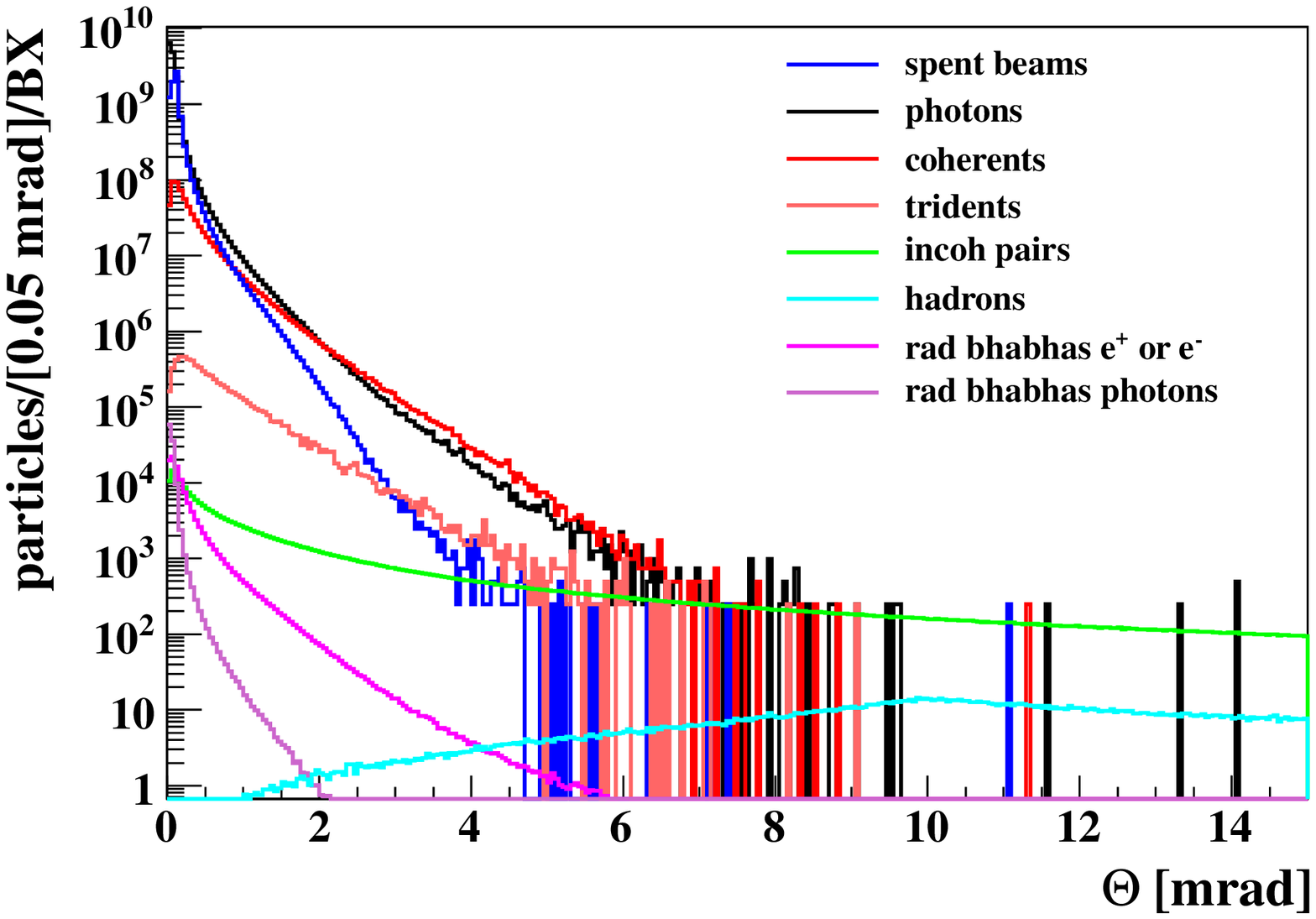}}\vspace{-0.2cm}
\end{minipage}
\caption{Energy distribution (left) and angular distribution (right) of the particles produced in beam-beam 
  background.}\label{Fig:2}
\end{figure}

Due to the strong focusing forces generated by the electromagnetic field during 
interaction, quite a lot of energy goes in the emission of synchrotron 
radiation photons, so called beamstrahlung photons, generating the long 
energy tail in the spent beam distribution.
On average two beamstrahlung photons are emitted per beam particle. Their energy
distribution is peaked at low values but a significant number of them can
reach the nominal beam energy, as shown in Fig.~\ref{Fig:2}(left). At
CLIC energies, where the beamstrahlung parameter $\Upsilon$ can be much larger
than 0.5, the emitted photons can turn into e$^+$e$^-$ pairs by interacting with the collective field of the oncoming beam, so called coherent process.
The energy spectrum of the produced pair depends on the beamstrahlung parameter, 
very low energy pairs are created only for high value of $\Upsilon$.
The angular distribution is boosted in the direction of the mother particle 
of the intermediate photon. The red lines in Fig.~\ref{Fig:2}(left) and 
Fig.~\ref{Fig:2}(right) show their expected energy and angular distribution.
In CLIC the coherent pair creation is the dominant process which produces 
e$^+$e$^-$ pairs during collision, 6.6$\times$10$^8$ coherent pairs are 
expected. Nevertheless at quantum beamstrahlung regime $\Upsilon > $1 and 
for very short bunch length the creation of pair may occur by an intermediate
virtual photon, in which case the pair production is said to occur by the 
trident process. The recent implementation of this process in GUINEA-PIG++ is 
described in~\cite{Jakob}, their production in the code
follow the one of the coherent pairs, except for the virtuality of the 
intermediate photon. The expected energy spectrum and their angular 
distribution for the nominal CLIC beams are shown by the light-red in 
Fig.~\ref{Fig:2}(left) and Fig.~\ref{Fig:2}(right), respectively. As coherent pairs
they follow mainly the beam direction while leaving the interaction region.
Their angular distribution is well confined in the 10 $mrad$ opening angle of
the interaction region beam pipe.

Most of the low energy e$^+$e$^-$ pairs are created at the future linear 
colliders by individual scattering of particles according to three main 
processes, the so called Breit-Wheeler ($\gamma \gamma \to $ e$^+$e$^-$), 
Bethe-Heitler (e$^\pm$ $\gamma$ $\to$ e$^\pm$ e$^+$e$^-$) and Landau-Lifshitz 
(e$^+$e$^-$ $\to$ e$^+$e$^-$e$^+$e$^-$) processes. 
Their are well known QED processes widely described in standard textbooks~\cite{qed}.
The main formulas implemented in GUINEA-PIG are described in~\cite{DanTh}. 
Their expected rate in CLIC is $\sim$ 330$\times$10$^3$, lower then the coherent 
pair one. Having very low energy, they can be highly deflected in the 
electromagnetic field of the incoming bunch therefore, they can enter in the 
detector region. The same process can lead to the production of muon pairs
as described in~\cite{Jakob}, the expected number of muons pair is 12.5 
per bunch crossing.

Hadronic events are produced at e$^+$e$^-$ colliders through the 
$\gamma \gamma \to$ hadrons reaction. The cross section is known experimentally
up to 200 GeV. Different parameterizations of the cross section with
the energy are implemented in GUINEA-PIG. According to the one in ~\cite{Hadr}
the expected number of $\gamma \gamma$ collisions per bunch crossing is
3.2 for a center of mass energy of the two photons of $>$ 2 GeV. 
The energy distribution of the produced hadrons is 
peaked at low energy and their angular distribution is more central then the
incoherent pairs one, allowing them to reach the central detector region.\\
     
\begin{wraptable}{l}{0.5\columnwidth} \vspace{-0.4cm}
\centerline{\begin{tabular}{|l|c|}
\hline
$\Delta$E/E$_{BS}$   &  29$\%$ \\ 
n$_{\gamma}$  & 2.1 per beam particle  \\ \hline
N$_{coherent}$ & 66$\times$10$^{7}$ \\
N$_{trident}$  & 67$\times$10$^5$ \\
N$_{incoherent}$ & 330$\times$10$^3$ \\
N$_{incoh-muons}$ & 12.50 \\
N$_{hadrons}$ & 3.2 \\
N$_{radiative-Bhabhas}$ & 110$\times$10$^3$ \\
\hline
\end{tabular}}\vspace{-0.3cm}
\caption{Average energy loss due to beamstrahlung and expected beam-beam 
         background rates per bunch crossing for the beam parameters reported 
	 in table~\ref{tab:1}.}
\label{tab:2}
\end{wraptable}\vspace{-0.4cm}

Radiative Bhabhas is another well known QED process, in which the binary 
collision of the electron-positron lead to the emission of a photon in the
final state (e$^+$e$^-$ $\to$ e$^+$e$^-$$\gamma$)~\cite{rbha}. At lowest order
the process (in t channel) can be modeled as a two steps reaction: 
first an e$^-$/e$^+$ is substituted by its photon equivalent spectrum,  
then the compton scattering of the photon on the e$^+$/e$^-$ is calculated.
The expected rate at CLIC is $\sim$110$\times$10$^3$. The energy 
and angular distributions of the scattered e$^-$/e$^+$ and photon are shown 
by the pink and light-pink curve in Fig.~\ref{Fig:2}(left) and Fig.~\ref{Fig:2}(right). 
Their energy is spread over a wide range (from 0 up to the nominal beam energy).
Their angular distributions are mainly peaked in the very forward direction.\\ 
All the expected beam-beam background rates we have studied 
are summarized in Table~\ref{tab:2}.  
The emittance values considered in the simulations include the budgets 
for imperfections. The actual values depend on the single machine and change 
during operation. 

\subsection{Machine imperfections and background rates}
\label{sec:figures}

If machine imperfections are well controlled the final emittance of the 
two beams can be lower then the one reported in Table~\ref{tab:1}, 
leading to a high luminosity and high background rate.
The overall correlation of the background rates with the horizontal
and vertical emittance of the two beams has been studied in~\cite{ipac10}.
In the following we report the evaluation of an upper limit of the rate of 
the two backgrounds of interest for the detectors, such as incoherent pairs 
and hadronic events. 
For this purpose we track the two beams in the Main LINAC and the BDS 
considering realistic imperfections and nominal beam parameters at the 
entrance of the LINAC, using the tracking code PLACET~\cite{plac}.
We consider here machine imperfections in the vertical plane only, which is the
most critical one due to the very small emittance. 
The vertical emittance at the entrance of the main LINAC is 10 nm and 
the machine imperfections
considered in the simulation are reported in Table~\ref{tab:3}.
 
\begin{table}[!h] \vspace{-0.3cm}
    \begin{center}
        \begin{tabular}{|c|c|c|}
          \hline
          \small imperfections   & \small dim.  & \small value \\
	  \hline 
          \small BPM vert. offset & \small $\mu$m &  14 \\
          \small BPM resolution   & \small $\mu$m &  0.1  \\ 
	  \small accelerating structure vert. offset & \small $\mu$m & 7 \\
	  \small accelerating structure vert. tilt & \small $\mu$rad & 142 \\
	  \small quadrupole vert. offset & \small $\mu$m & 17 \\
          \small quadrupole vert. roll & \small $\mu$rad & 100 \\
	  \hline
          \small beam parameters   & \small dim.  & \small value \\
	  \hline 
          \small Bunch charge N    & \small particles  & 3.72e+09 \\
          \small Bunch length $\sigma_{z}$  & \small $\mu$m  & 44 \\
          \small hor. emittance $\gamma\epsilon_{x}$  & \small nm  & 660 \\
          \small vert. emittance $\gamma\epsilon_{y}$  & \small nm  & 10 \\
	  \hline 
        \end{tabular}\vspace{-0.2cm}
    \caption{ \small Values of the machine imperfections and beam parameters 
    used in the main LINAC simulations.    
    \label{tab:3} }
    \end{center}
\end{table}\vspace{-0.3cm} 

These imperfections are enough to bring the vertical emittance of the 
nominal beams to growth up to several order of magnitude if no correction 
scheme is applied to the machines. 
When the Beam-Based-Alignment (BBA), described in~\cite{dpac09}, is applied to
the machines the average emittance growth at the end of the ML stays 
well below five nm, which is the budget for static imperfections in the main 
LINAC. We steer the beams coming from the corrected linacs into the BDS, and 
track them to the IP, without any imperfections in the BDS. 
The bunch shapes, so obtained, are used to compute luminosity and background 
rates again.
This procedure allow us to evaluate the effect of imperfections in the Main 
Linac only on the luminosity and on the background rates.
Moreover since further machine imperfections in the BDS 
would only lower the luminosity and background rates, this assumption ensure 
that we estimate a maximum value for the background rates.\\


\begin{figure}
\begin{minipage}[b]{7.1cm}
\centerline{\includegraphics[width=0.6\columnwidth,angle=270]{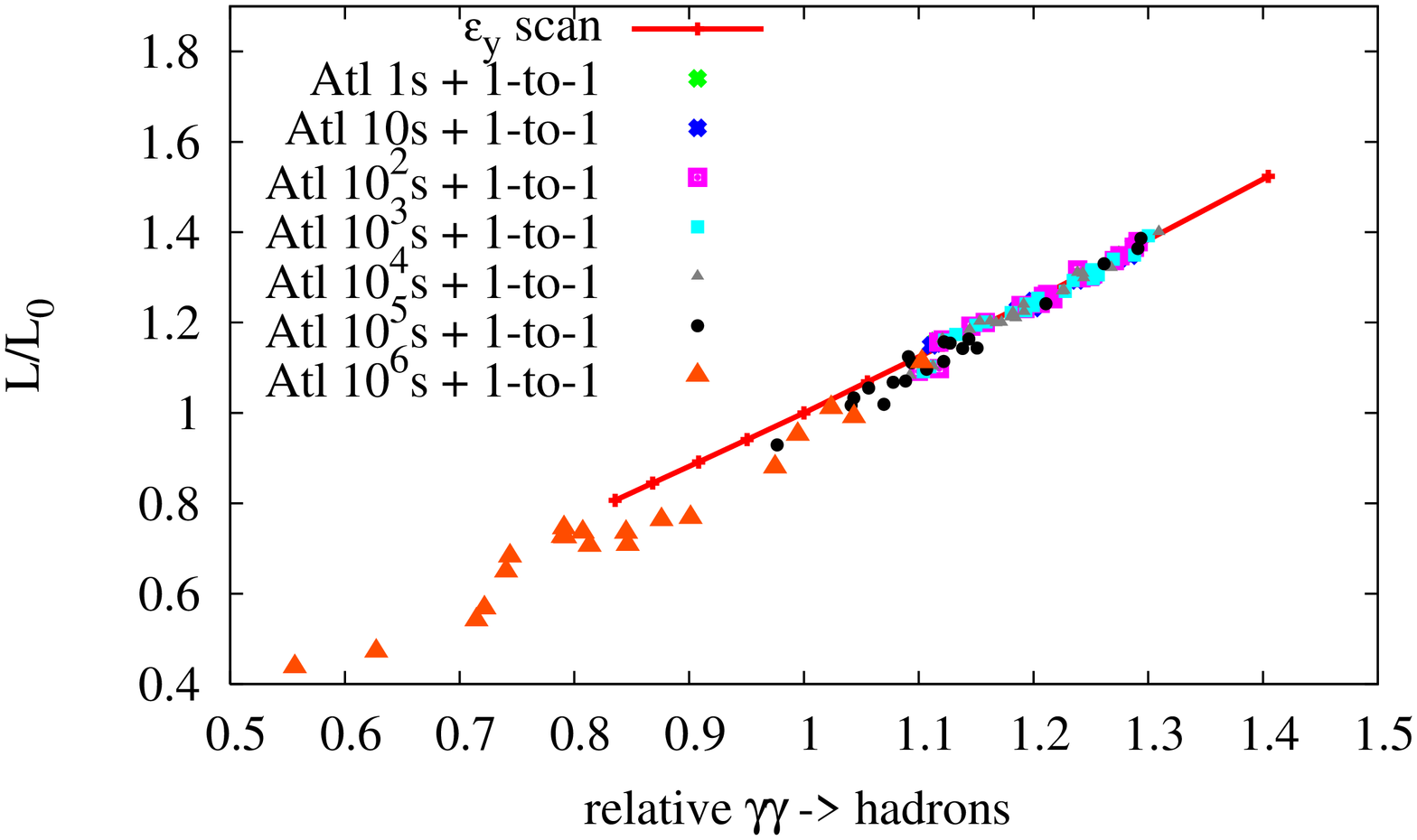}}\vspace{-0.3cm}
\end{minipage}
\begin{minipage}[b]{7.1cm}
\centerline{\includegraphics[width=0.6\columnwidth,angle=270]{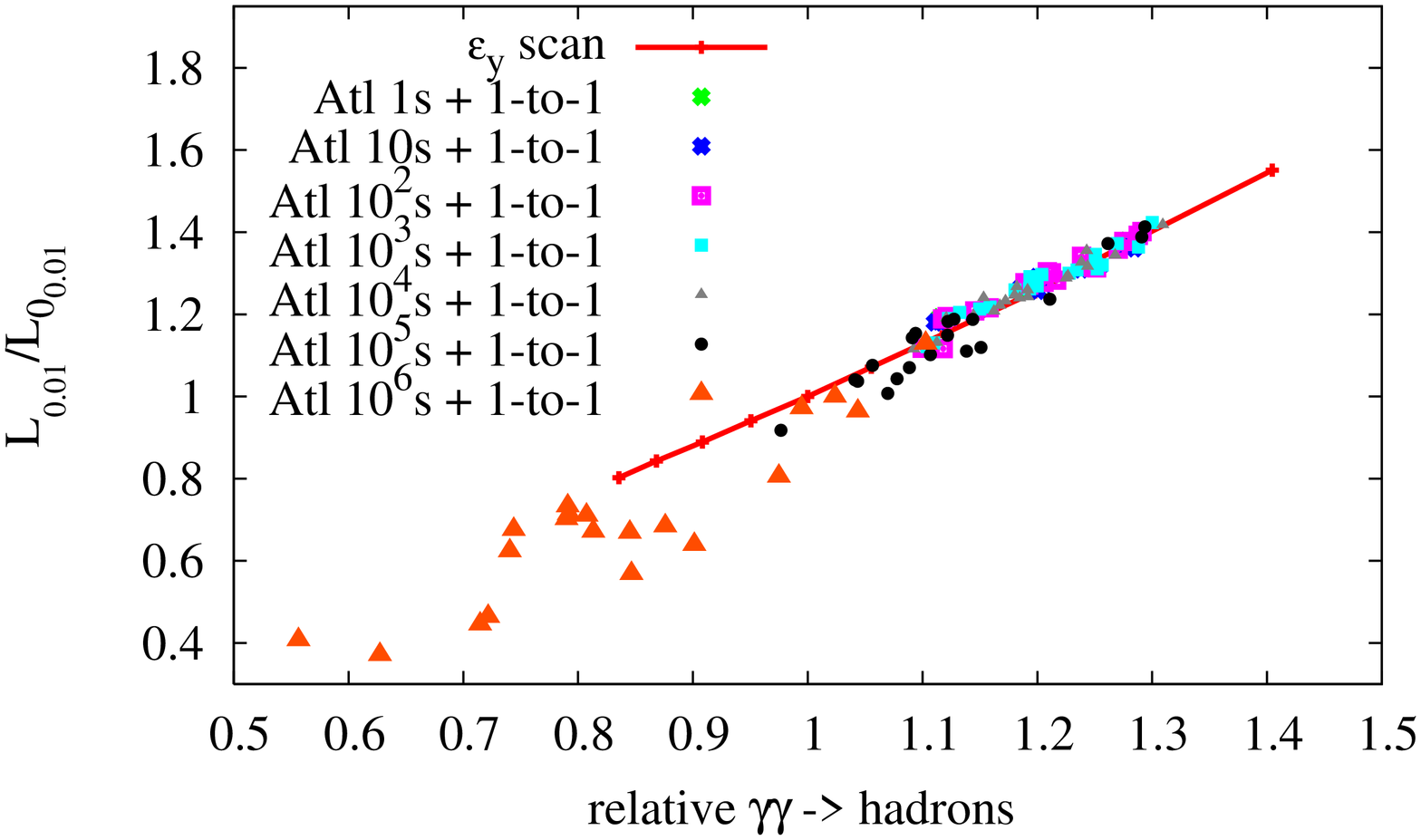}} \vspace{-0.3cm}
\end{minipage}
\caption{ Total luminosity L(left) and Peak luminosity L$_{0.01}$(right) vs number of 
    $\gamma\gamma$ $\to$ hadrons events, normalized to the nominal values, 
    for perfect machines and different vertical emittance 
    values, and for corrected machines and nominal vertical emittance.}
    \label{Fig:3}
\end{figure}

\begin{figure}
\begin{minipage}[b]{7.1cm}
\centerline{\includegraphics[width=0.6\columnwidth,angle=270]{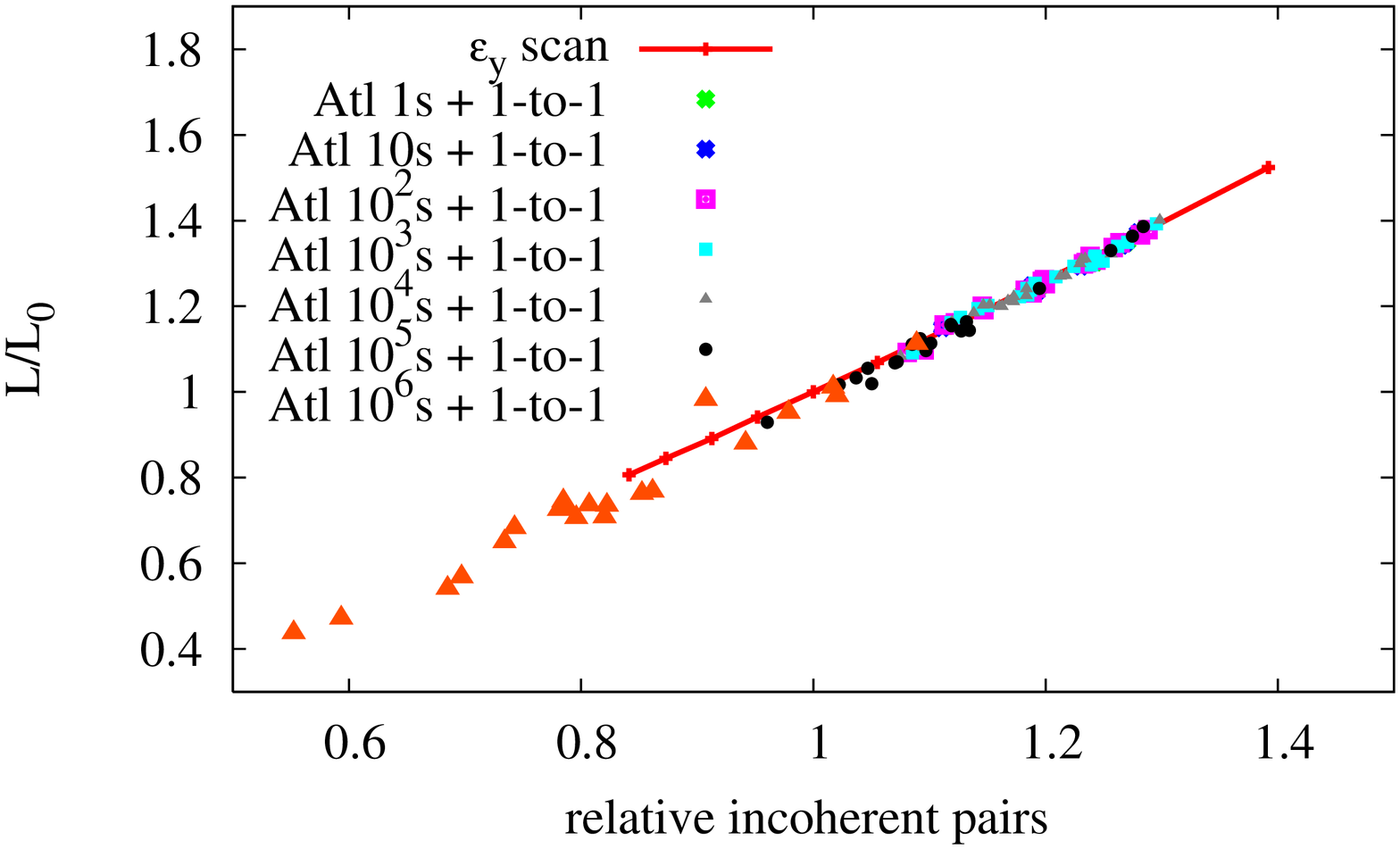}}
\end{minipage}
\begin{minipage}[b]{7.1cm}
\centerline{\includegraphics[width=0.6\columnwidth,angle=270]{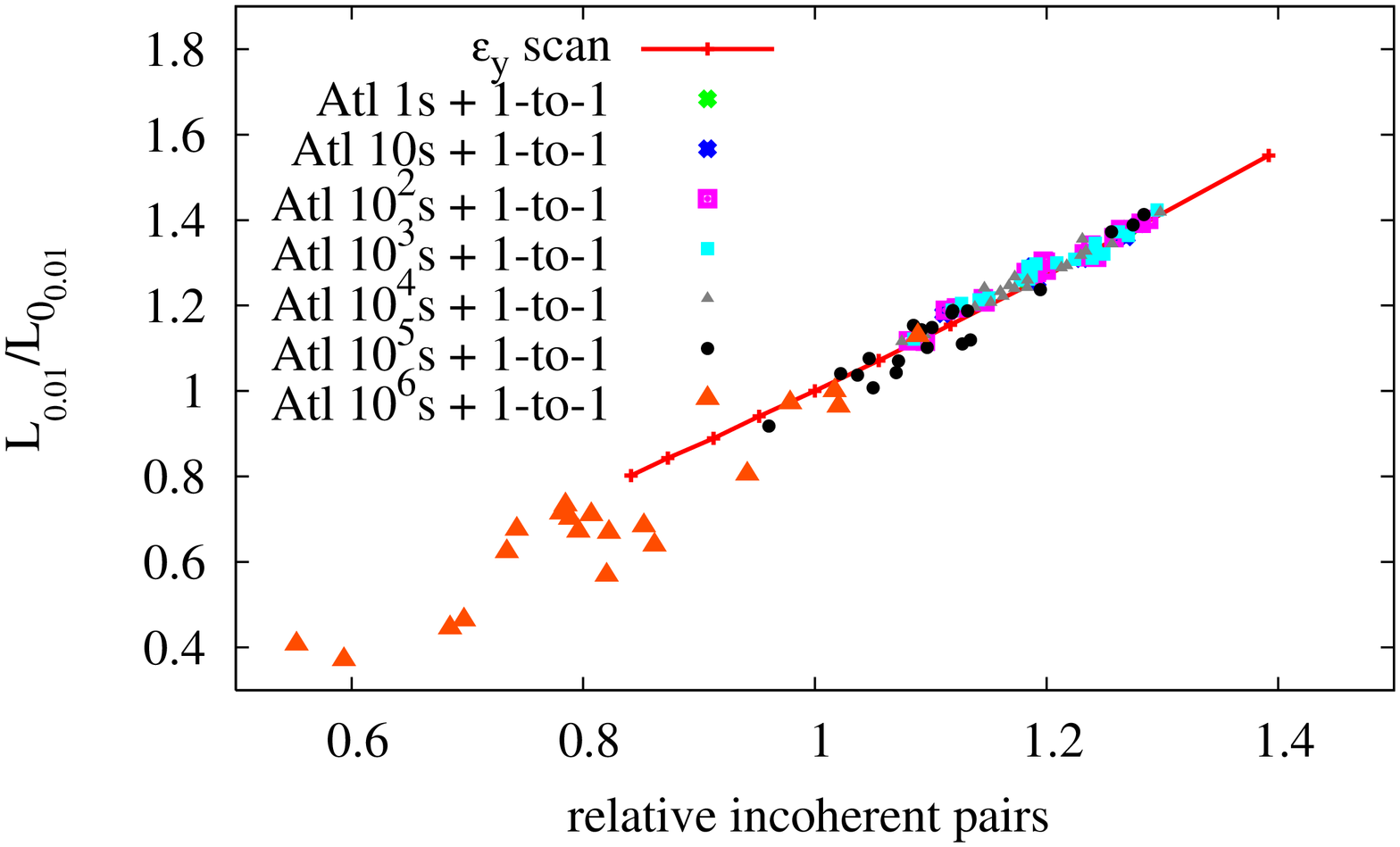}}
\end{minipage}
\caption{ Total luminosity L(left) and Peak luminosity L$_{0.01}$(right) vs number of 
  incoherent pairs events, normalized to the nominal values, for perfect machines 
  and different vertical emittance values, and for corrected machines and nominal 
  vertical emittance.}\label{Fig:4}
\end{figure}

The cases of perfect machines and scaled beam emittances, 
as in Figs.~\ref{Fig:3} and ~\ref{Fig:4} 
are compared with the cases of corrected machines and initial nominal beam 
parameters.
The Total and Peak luminosities are shown in separeted plots. We define Peak
Luminosity the luminosity contained into $\pm$1$\%$ around the 3 TeV CM peak.\\
The correlation between luminosity and background rates variations is linear 
for both the background types considered, and it stays linear for 
both total(L) and peak(L$_{0,01}$) luminosity.
The fluctuation in the variation of luminosity and background rates 
for the corrected machines is on average $\sim$5$\%$,
indicating that different emittance values can be reached by the different 
machines after the linac BBA correction. In these simulations 
the mean luminosity reached by the corrected machines is about 30$\%$ higher than the values reported in Table~\ref{tab:1}, and on average 25$\%$ more background with respect to the values in Table~\ref{tab:2}. 
When 10$^{6}$ s of ground motion is applied to the Main LINAC the
luminosity and background rates start to fluctuate more around the 
linear behavior (the fluctuations of the background rates are $>$ 15$\%$). 
The majority of the machines have a relative low luminosity while still a 
significant number of hadronic events and incoherent pairs can be produced.
Even though the background rates stay below the values quoted
in Table~\ref{tab:2}. 
A safety margin of 50$\%$ more luminosity and 40$\%$ more background (for 
both hadronic events and incoherent pairs) can be defined. Detectors should
thus be able to handle this level in order not to have to reduce
luminosity to reduce background. 

\section{Synchrotron Radiation photons from the final doublet}

In order to provide an acceptable cleaning efficiency of the desired beam halo
the collimation apertures are determined from the following conditions:

\begin{enumerate}
\item  minimize the synchrotron radiation photons emitted in 
      the first final quadrupole magnet which can hit the 2nd final 
      quadrupole (QD0);
\item  minimize the beam particles that can hit either QF1 or QD0.
\end{enumerate}

Macroparticles with high transverse amplitude are tracked using the code 
PLACET~\cite{plac}, taking into account the emission of synchrotron radiation
and all the non linear elements of the system. The particles positions and 
angles have been checked at the entrance, in the middle and at the exit of QF1
and QD0. The dangerous particles are efficiently removed for collimator 
apertures of 
$<$ 15 $\sigma_{x}$ in the horizontal plane and of $<$ 55 $\sigma_{y}$ in the 
vertical plane. Therefore we define 15 $\sigma_{x}$ and 55 $\sigma_{y}$ 
as the collimation depths~\cite{resta}.\\

\begin{figure}[h] 
\centerline{\includegraphics[width=1.0\columnwidth]{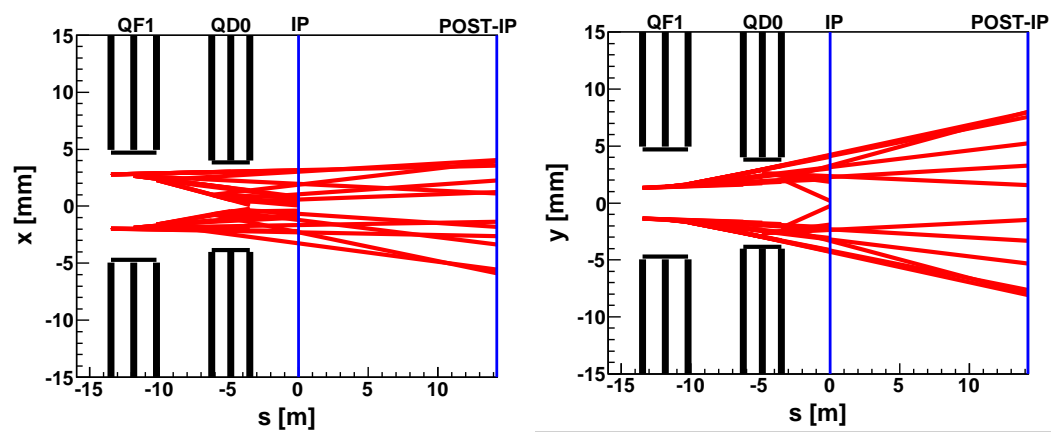}}\vspace{-0.2cm}
\caption{Incoherent Synchrotron radiation photon fans at 3 TeV.}\label{Fig:5}
\end{figure}

Fig.~\ref{Fig:5} shows the residual synchrotron radiation fans from the final
quadrupoles QF1 and QD0 to the IP for an envelop covering 15 and 55 standard 
deviations in $x$ and in $y$, respectively. At the IP the photon cone is 
inside a cylinder of radius of five mm, which is well inside the beam pipe 
aperture. Therefore, in principle, they are not an issue of concern for the 
detectors.\\

\newpage

\begin{wrapfigure}{r}{0.5\columnwidth} \vspace{-0.5cm}
\centerline{\includegraphics[width=0.45\columnwidth]{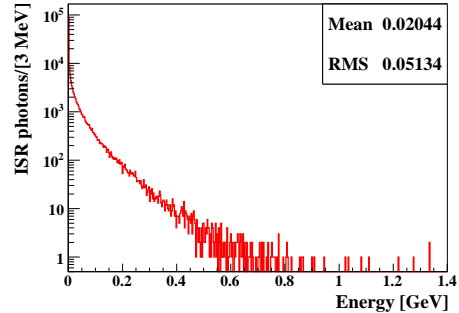}}\vspace{-0.2cm}
\caption{Energy of synchrotron radiation photons emitted in the final 
		doublet.}\label{Fig:6}
\end{wrapfigure} \vspace{-0.5cm}

The distribution of the expected energy of the radiated photons in the final 
doublet for a perfect machine and nominal beam parameters is shown in 
Fig.~\ref{Fig:6}. The spectrum is peaked at very low energy, i.e. $<$ 1 MeV, 
with an energy tail up to $\sim$ 1 GeV.
The number of radiated photons is $\sim$ 1 for beam particle.
An internal shielding of the beam pipe in the final doublet magnets region 
should be foreseen.\\

\section{Conclusion}

Beam-beam effects at CLIC 3 TeV CM energy have been reviewed. The expected
production rates, their energy and angular distribution evaluated. A safety
margin of 40$\%$ of the background processes of interest for the detectors 
is estimated. The ISR photon fans coming from the final doublet is shown 
to be well inside the beam pipe aperture at the IP, considering the nominal
collimation depths. Their energy distribution is peaked at few MeV.
 

\begin{footnotesize}


\end{footnotesize}

\end{document}